\documentclass[prb,twocolumn,superscriptaddress]{revtex4-1}

\usepackage{amsmath}
\usepackage{amssymb}
\usepackage{graphicx}
\usepackage[caption=false]{subfig}
\usepackage{braket}
\usepackage{amssymb}
\usepackage{natbib,hyperref}
\usepackage{color}

\begin{document}

\title{Odd-frequency Pairing in the Edge States of Superconducting Pnictides in the Coexistence Phase with Antiferromagnetism}

\author{Cody Youmans}
\affiliation{Physics Department, City College of the City University of New York, New York, New York 10031, USA}
\affiliation{The Graduate Center of the City University of New York, New York, New York 10016, USA}
\author{Areg Ghazaryan}
\affiliation{Physics Department, City College of the City University of New York, New York, New York 10031, USA}
\author{Mehdi Kargarian}
\affiliation{Department of Physics, Condensed Matter Theory Center and Joint Quantum Institute, University of Maryland, College Park, MD 20742, USA}
\author{Pouyan Ghaemi}
\affiliation{Physics Department, City College of the City University of New York, New York, New York 10031, USA}
\affiliation{The Graduate Center of the City University of New York, New York, New York 10016, USA}

\date{\today}

\begin{abstract}
In some of the Ferro-pnictide materials, spin density wave order coexists with superconductivity over a range of doping and temperature. In this paper, we show that odd-frequency pairing emerges on the edges of pnictides in such a coexistence phase. In particular, the breaking of spin-rotation symmetry by spin density wave and translation symmetry by the edge can lead to the development of odd-frequency spin-triplet Cooper pairing. In this case, the odd-frequency pairing has even parity components, which are immune to disorder. Our results show that pnictides are a natural platform to realize odd frequency superconductivity, which has been mainly searched for in heterostructures of magnetic and superconducting materials. The emergence of odd-frequency pairing on the edges and in the defects can be potentially detected in magnetic response measurements.

\end{abstract}

\maketitle
\section{Introduction}
The discovery of superconductivity at elevated temperatures in pnictides revived interest in the study of high-temperature superconductivity\cite{Johnston2010,Paglione2010,Stewart2011,Chubukov2012,Fernandes2017,Kamihara2008,Rotter2008}. 
Similar to cuprates\cite{Bednorz1986}, pnictides present different phases and phase transitions which can be tuned by changing temperature and doping level. Contrary to cuprates, which are antiferromagnetic Mott insulators at low doping, the parent compounds of pnictides are metallic and develop spin-density wave (SDW) order which takes the form of ferromagnetic stripes aligned antiferromagnetically (i.e., $(\pi,0)$ or $(0,\pi)$ SDW)\cite{Dai2015}. In both cuprates and pnictides, upon doping, the magnetic order is suppressed and superconductivity emerges. It is widely believed that superconducting gap in pnictides is of extended $s$-wave ($s_\pm$)  type\cite{Hirschfeld2011}, which changes sign between different Fermi pockets. Contrary to cuprates, where superconductivity develops after antiferromagnetism disappears, in some families of pnictides, over a certain range of dopings and temperatures, superconductivity and SDW coexist\cite{Wen2011,Pratt2009,Laplace2009,Chen2009}. 

The presence of edge states has long been used to identify the structure of the superconducting pairing gaps in unconventional superconductors. A prominent example is the appearance of Andreev bound states (ABS) on the (110) edge of cuprates, resulting from the d-wave structure of the superconducting gap\cite{PhysRevLett.72.1526}. The signature of such edge states has been observed in tunneling experiments \cite{GREENE2003162}. Emergence of ABS at the edges of pnictides has also been proposed as a signature of the extended s-wave superconductivity\cite{Ghaemi2009,Araujo2009,Huang2010,Lau2014}. On another front, it has been argued that in the normal phase, the non-trivial topological character of the electronic band structure of pnictides can lead to the development of edge states which are spin-degenerate in the paramagnetic phase and split into spin-polarized edge bands in the SDW phase\cite{Lau2013}. In this paper we consider the coexistence of SDW and extended s-wave superconductivity and show that further unconventional types of superconducting pairing, known as odd-frequency pairing, should develop on the edges. 





Starting from the work of Berezinskii \cite{Berezinskii1974}, new types of superconducting pairings, known as odd-frequency pairings, have been postulated. Cooper pair wave-functions satisfy the fermionic antisymmetry requirement typically through the spin (e.g., s-wave spin-singlet pairing) or internal angular momentum of the pair (e.g., p-wave spin-triplet pairing). Odd-frequency Cooper pairs satisfy the antisymmetry requirement in terms of the Matsubara frequency or relative time coordinate of the paired electrons\cite{Tanaka2012,Linder2017,Abrahams1995,Dahal2009}. In this regard, four types of pairing symmetries are of immediate interest: odd-frequency spin-singlet odd-parity (OSO), odd-frequency spin-triplet even-parity (OTE), even-frequency spin-singlet even-parity (ESE), and even-frequency spin-triplet odd-parity (ETO). The familiar s-wave spin-singlet and p-wave spin-triplet superconducting phases belong to the ESE and ETO classes, respectively\cite{Tanaka2012}. 
 Identification of the mechanisms for generating robust odd-frequency pairing in disordered Fermi systems\cite{Kirkpatrick1991,Belitz1992,Belitz1999} showed the potential of experimentally realizing this novel superconducting state.
 
A promising platform to realize odd-frequency pairing is the heterostructure of a superconductor and a ferromagnetic metal\cite{Bergeret2001,Bergeret2005,Tanaka2005,Buzdin2005,Eschrig2007,Eschrig2015}. Simply speaking, the ferromagnetic layer lifts the spin degeneracy, leading to a mixture of spin-singlet and spin-triplet Cooper pairs. In addition, the breaking of translational symmetry due to the interface mixes even-parity and odd-parity pairings. Consequently, in compliance with fermionic anti-commutation rules, odd-frequency triplet pairing can be generated at an interface between a superconductor and a ferromagnet.  Recent advancements in building heterostructures of materials with magnetic and superconducting properties bring the experimental realization of odd-frequency pairing well within reach. It has also been shown that that spin-orbit coupling (SOC) can also lead to odd-frequency triplet pairing in superconducting materials with translational symmetry breaking in the absence of magnetism\cite{Cayao2017}.




Various experimental signatures of odd frequency pairing have been proposed\cite{Yokoyama2007,Komendova2015,DiBernardo2015,Fleckenstein2018}. For a diffusive metal in proximity to an ETO superconductor, proximity induced OTE pairing due to translational symmetry breaking is directly linked with a zero energy peak in the density of states (DOS) within the superconducting gap induced in the metal\cite{Tanaka2007Jan}. Odd-frequency pairing is also credited for its potential to elicit a paramagnetic Meissner response \cite{Yokoyama2011,Lee2017,DiBernardo2015Meissner}. Another interesting link has been suggested between odd-frequency superconductivity and Majorana bound states, through a proportionality relation between the local DOS for a zero-energy state and the odd-frequency pairing amplitude\cite{Asano2013,Daino2012,Cayao2017}.

In this paper we explore pnictides with coexisting SDW and superconductivity to realize odd-frequency pairing. {For pnictides in the coexistence regime, SDW breaks spin-rotation symmetry, while translational symmetry is also broken at the edge of the sample. It is thus conceivable that both types of odd-frequency pairing (OSO and OTE) would  be generated at the edge of the sample. The generation of OTE is particularly important, because such superconducting pairing is robust in the diffusive regime \cite{Tanaka2007Jan,Tanaka2012}. Therefore, pnictides provide a natural platform to explore odd-frequency pairings without the need for considering complex heterostructures. Also, while this work focuses  on odd-frequency pairings at the edges, our results suggest the possibility of realizing such pairings at SDW defects. It has been predicted that odd-frequency pairing will lead to anomalous magnetic responses\cite{Fominov2015,Asano2014}. Such effects could potentially explain the enhancement of superfluid density along SDW defects as observed in local magnetic response measurements\cite{Kalisky2010}. This connection will be the subject of our future efforts.}

This paper is organized as follows. In Sec.~\ref{OrbitalModel} the two orbital model of iron pnictides is reviewed and the edge states for the paramagnetic phase (without SDW) are discussed. Sec.~\ref{SDWPhase} is devoted to the inclusion of the SDW term in the non-superconducting mean-field Hamiltonian. In particular, it addresses the doubling of the unit cell and Fermi surface reconstruction due to the SDW. The coexistence regime of SDW and superconductivity is discussed in Sec.~\ref{CoexistencePhase}. In Sec.~\ref{OddFrequency} the odd frequency  correlators are evaluated and their physical implications are discussed. Due to the fact that the two-orbital model does not capture the correct lattice symmetries in pnictides, in Sec.~\ref{fiveo} we study the emergent odd-frequency superconducting states in the five orbital model using exact diagonalization. In this section we also include the spin-orbit (SOC) coupling and show that SOC would lead to the generation of odd frequency triplet pairings which are not present in the absence of SOC. Finally, Sec.~\ref{Conclusion} presents conclusions, proposal for experimental verification and directions for future research.

\section{Two orbital model}
\label{OrbitalModel}
A minimal model for pnictides consists of a square lattice of iron atoms. While five orbitals on each iron site are necessary for a comprehensive microscopic band description of pnictides\cite{KurokiTanaka2008}, many of their properties can be understood using a two orbital effective model \cite{Fernandes2017,Raghu2008,Ran2009} including the $d_{xz}$ and $d_{yz}$ orbitals of iron,
\begin{equation}   
\label{OrbHam}
\mathcal{H_{\lambda}}(\vec{k})=\phi_0(\vec{k})\lambda^0 +\phi_1(\vec{k})\lambda^1 +\phi_3(\vec{k})\lambda^3.
\end{equation}

\begin{figure}[h]
\centering
\subfloat[]{{\includegraphics[trim={0cm .57cm 0cm 0cm},clip,scale=.33]{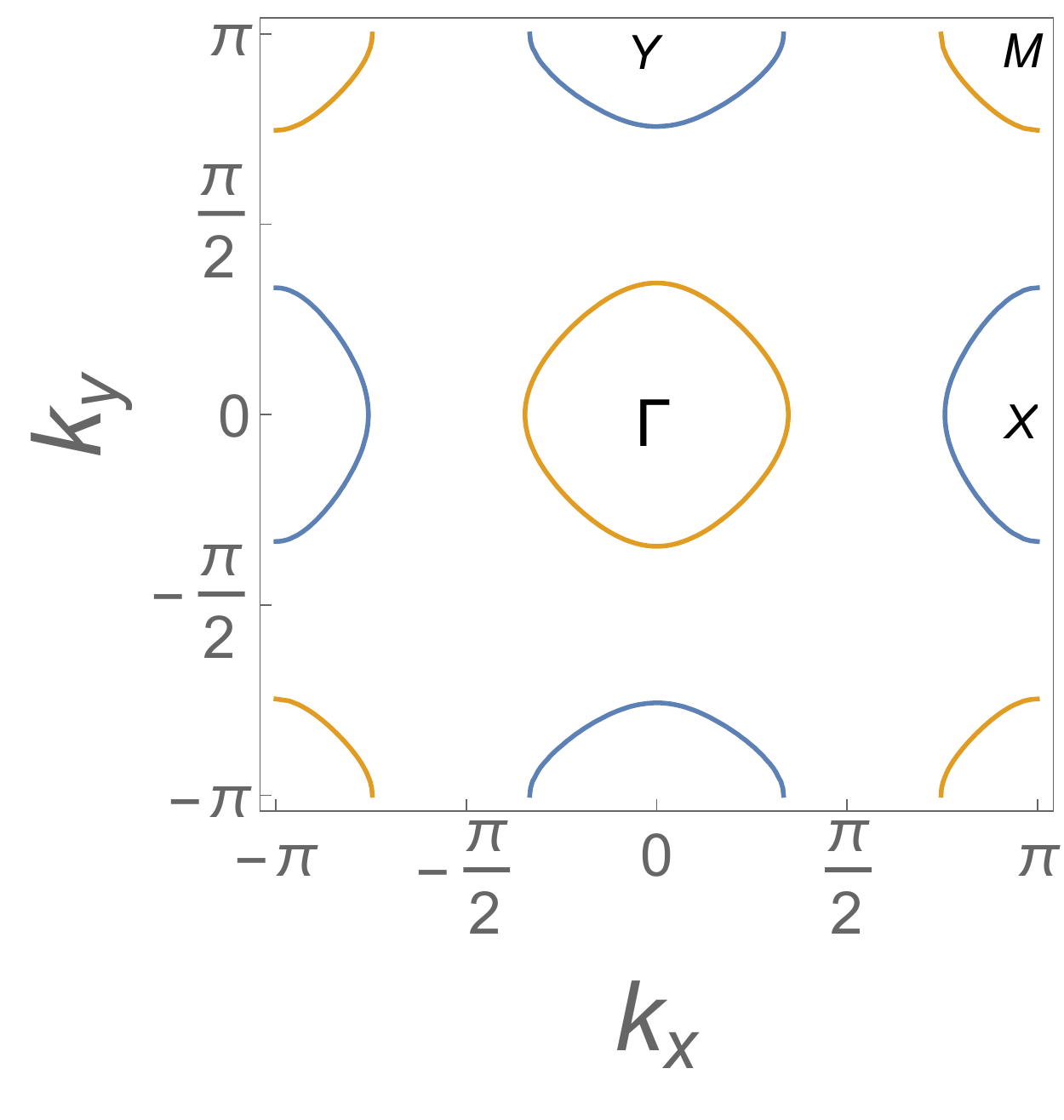}  }}
    \subfloat[]{{\includegraphics[trim={5.8cm 9.23cm 4cm 8.9cm},clip,scale=.44]{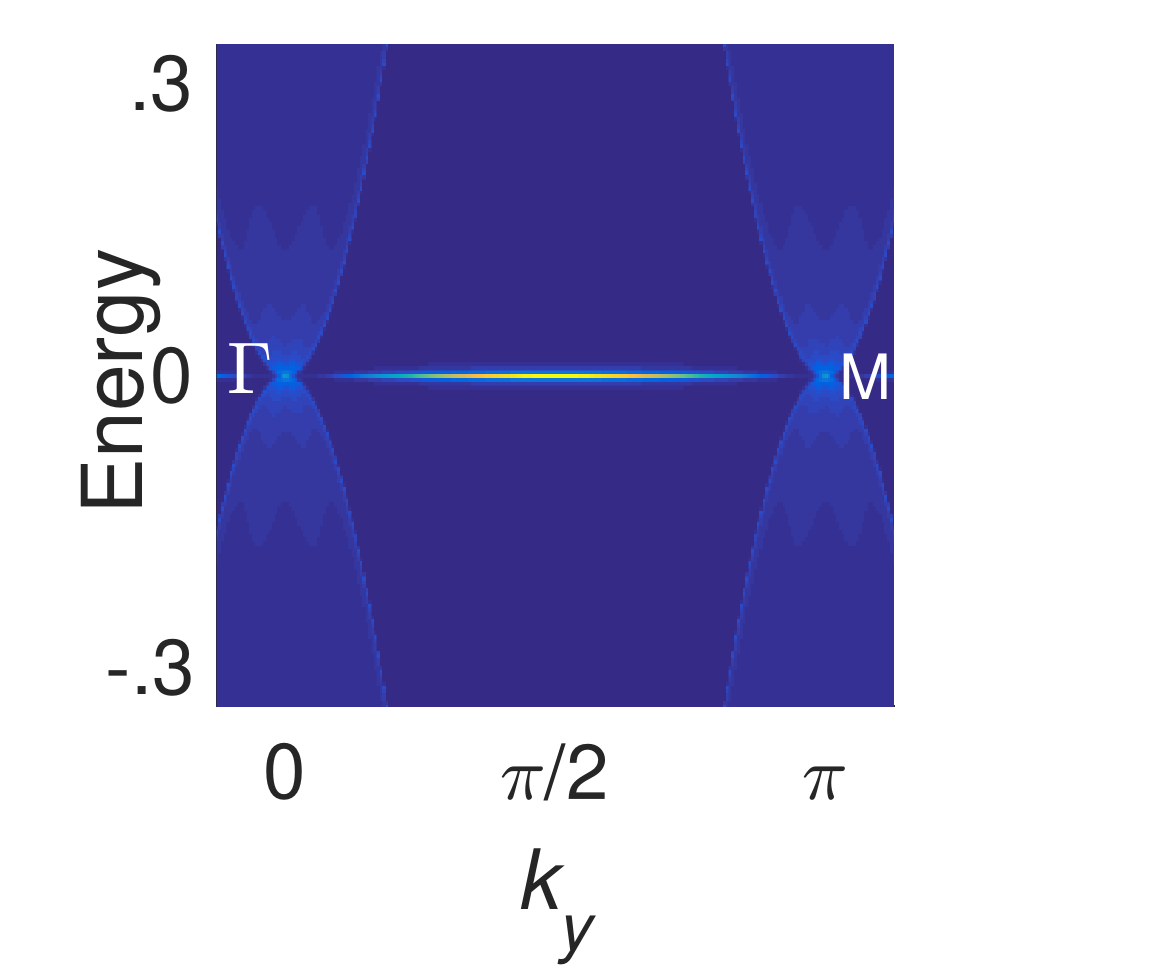}  }}
\caption{(a) Electron (blue) and hole (orange) orbital Fermi pockets in first BZ, with high symmetry points $\Gamma$, X, Y, and M labeled. (b) Flat-band on edge of topologically equivalent semimetal (\ref{eq:SSH}), connecting projected bulk QBTs at $\Gamma$ and $M$ points.}
\label{fig:BZandflatband}
\end{figure}

Here, $\phi_0(\vec{k}) = 2(t_2 + t_2^\prime)\cos{k_x}\cos{k_y} + 2t_1^\prime(\cos{k_x} + \cos{k_y}) - \mu_F,  \ \phi_1(\vec{k}) = 2(t_2-t_2^\prime)\sin{k_x}\sin{k_y}$, and $\phi_3(\vec{k}) = 2t_1(\cos{k_x}-\cos{k_y})$ are defined on the 2D Brillouin zone (BZ), and $\mu_F$ is the chemical potential. In the Hamiltonian (\ref{OrbHam}), $\lambda^0$ and $\lambda^i$ (i=1,2,3) are the identity and Pauli matrices acting on the orbital basis $\Psi_{\lambda,\vec{k}} =\left(d_{xz,\vec{k}},d_{yz,\vec{k}}\right)^T$. $t_1$($t_1'$) and $t_2$($t_2'$) are nearest and next nearest neighbor hoping parameters between different orbitals. For the two-band model we set $t_1=1$, $t_2=1.7$, $t_1^\prime=.2$, $t_2^\prime=0$, and $\mu_F=1.1$. With these parameters, the Fermi pockets of the electron (hole) bands $\mathcal{E}_{e(h)}=\phi_0+(-)\sqrt{\phi_1^2+\phi_3^2}$ are shown in Fig. \ref{fig:BZandflatband}(a). Throughout the paper all energies are in units of $t_1$.



In the superconducting phase, the Bogoliubov-De Gennes (BdG) Hamiltonian reads, 
\begin{equation} \mathcal{H_{\mu\sigma\lambda}}(\vec{k}) =\mu^3\sigma^0\mathcal{H_{\lambda}}(\vec{k}) + \Delta_s^\pm(\vec{k}) \mu^2\sigma^2\lambda^0,
\end{equation}
where $\Delta_s^\pm(\vec{k}) = \Delta\cos{k_x}\cos{k_y}$ is the extended s-wave superconducting gap. $\mu$ and $\sigma$ are Pauli matrices acting on superconducting particle-hole and physical spin, resp.. 



For an edge along $\hat{y}$, momentum $k_y$ parallel to the edge is conserved and the Hamiltonian $\mathcal{H}_\lambda(-i\partial_x,k_y)$ for fixed $k_y$ describes an effective 1D system extending along $\hat{x}$. Exact diagonalization of these 1D Hamiltonians, shows the presence of subgap surface-bands. Development of these edge states can be also analytically understood by putting $\phi_0=0$ in Hamiltonian \ref{OrbHam}. For such a Hamiltonian, the hole pockets encircling $\Gamma = (0,0)$ and $M=(\pi,\pi)$ in figure \ref{fig:BZandflatband}(a) turn into points which correspond to quadratic band touchings (QBTs) (see figure \ref{fig:BZandflatband}(b)). In this case, $k_y$ labels gapped 1D Hamiltonians extending perpendicular to the edge:
\begin{equation}
\label{eq:SSH}
\tilde{\mathcal{H}}_\lambda(-i\partial_x,k_y)\approx v {i\partial_x}\lambda^1 +(\alpha \partial_x^2-\beta)\lambda^3,
\end{equation}
where the parameters $v$, $\alpha$, and $\beta$ depend on $k_y$.
For each $k_y \notin \{ 0,\pm\pi \}$, the spectrum of Hamiltonian \ref{eq:SSH} contains a midgap edge state\cite{ssh} (Fig.~\ref{fig:BZandflatband}(b)). The deformed Hamiltonian with $\phi_0=0$ would thus exhibit 1D flat-band edge states connecting the projected bulk QBT points.

Restoring $\phi_0$ introduces a dispersion to the flat-band edge states, which remain within the bulk gap\cite{Lau2013}. With the onset of superconductivity, these edge-bands disperse into the ABS\cite{Lau2014} within the superconducting gap formed at the Fermi pockets. The ABS states result from the extended s-wave structure of the superconducting gap\cite{Ghaemi2009}. While remaining nodeless on each Fermi pocket, the extended s-wave superconducting gap $\Delta_s^\pm(\vec{k})$ changes sign between pockets. Andreev reflection of low-energy electronic quasiparticles at the edge of the sample involves scattering between bulk Fermi surfaces with opposite sign of the superconducting gap, resulting in ABS \cite{Ghaemi2009,Huang2010}.

\section{Spin-Density-Wave Phase}
\label{SDWPhase}

{The coexistence of SDW order and extended s-wave superconductivity was a theoretically challenging phenomenon when initially realized experimentally \cite{Wen2011, Kalisky2010, Laplace2009, Pratt2009, Chen2009}. The puzzling feature of this phase is mainly due to the fact that $\vec{Q}=(\pi,0)$ SDW order mixes the Fermi pockets with opposite sign of superconducting gaps, which might appear to suppress superconductivity. It was soon theoretically shown that this expectation is not correct and in fact SDW and superconductivity do not compete with each other\cite{Parker2009,Hinojosa2014,Dai2015, Lau2013,Ghaemi2011,Ran2009}.}

\begin{figure}[h]%
    \centering
   \subfloat[]{{\includegraphics[trim={5.3cm 1.5cm 9cm 17.1cm},clip,
   scale=.45]
   {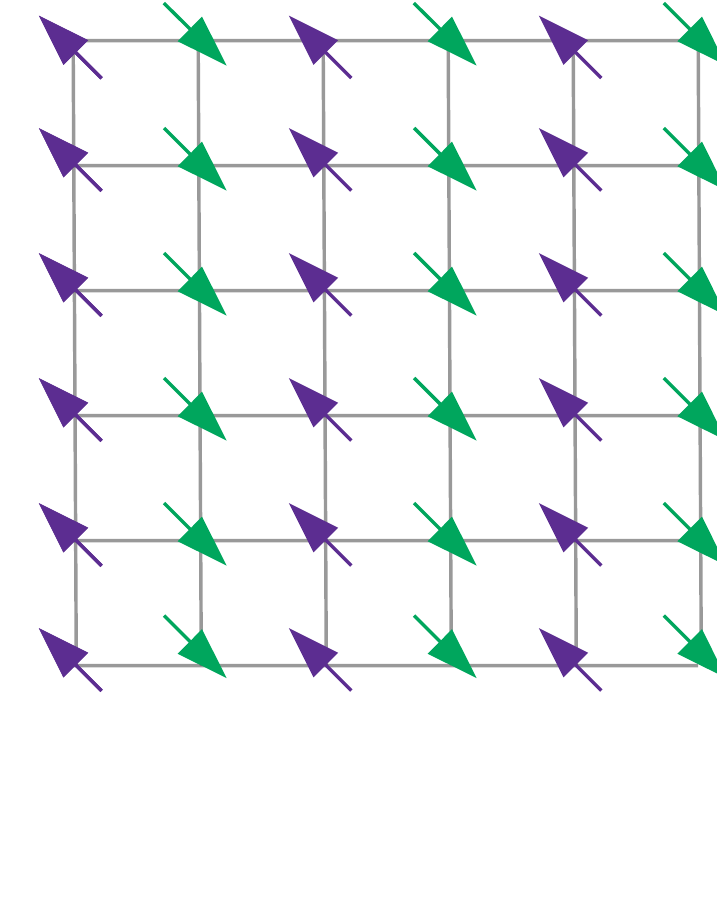} }}%
    \qquad
    \subfloat[]{{\includegraphics[
    scale=.35]
    {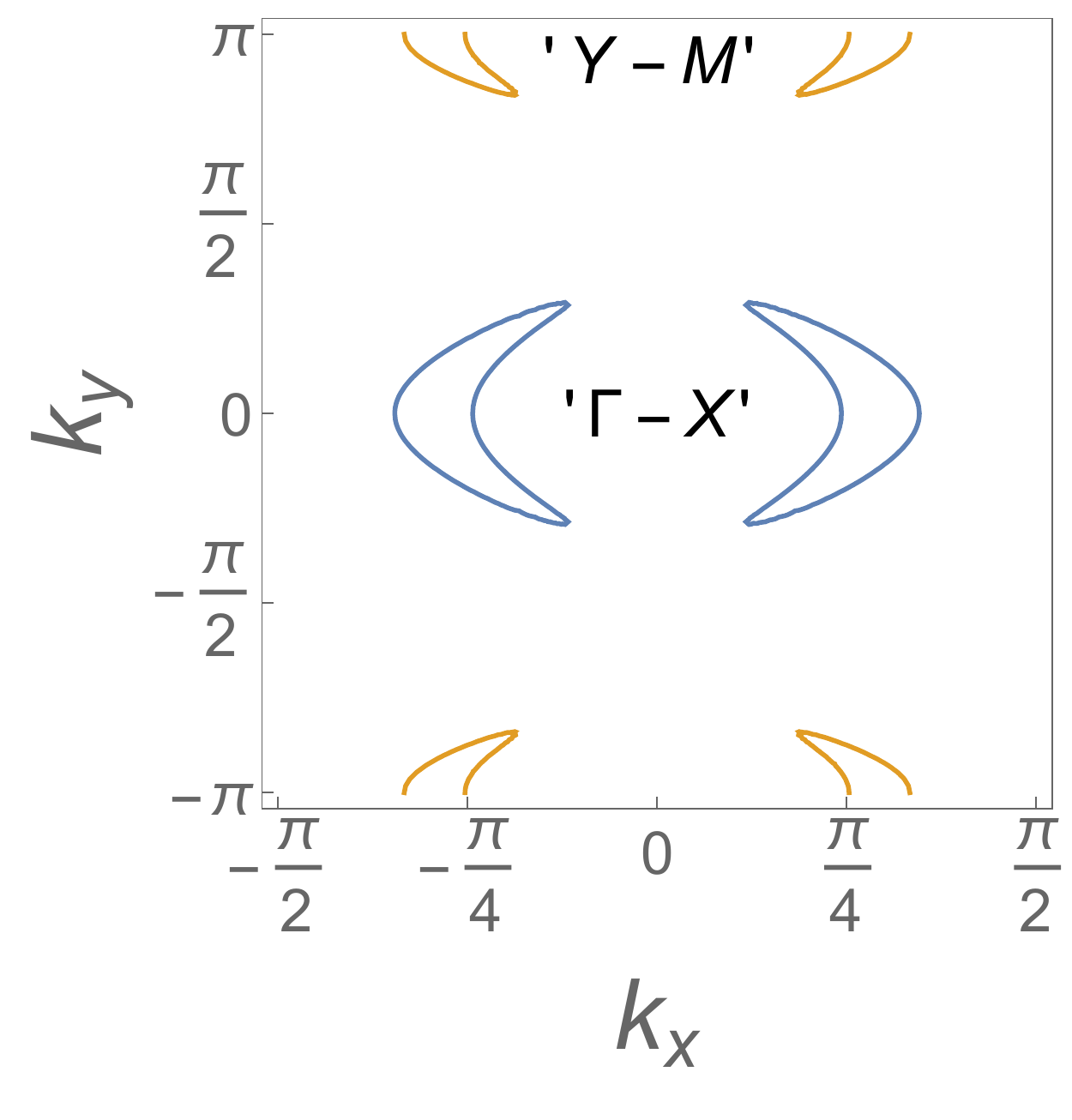} }}%
\caption{(a) $\vec{Q}=(\pi,0)$ spin-density wave in pnictides. (b) Folded BZ resulting from $\vec{Q}=(\pi,0)$ order.}
\label{fig:SDW}
\end{figure}

SDW order with wave vector $\vec{Q}=(\pi,0)$  doubles the unit cell along the x-direction (Fig.~\ref{fig:SDW}). 
In momentum space, this can be captured via the Bloch wave function of the form $\Psi_{\tau\lambda}(\vec{k})= \begin{pmatrix} \Psi_{\lambda}(\vec{k}) \\ \Psi_{\lambda}(\vec{k}+ \vec{Q})\end{pmatrix}$ \cite{Ghaemi2011}. The bulk Hamiltonian in the folded BZ reads, 
\begin{equation} \label{hamiz}
\mathcal{H_{\sigma\tau\lambda}}(\vec{k})=\mathcal{H_{\lambda}}(\vec{k})\oplus\mathcal{H_{\lambda}}(\vec{k}+\vec{Q})+\Delta_{SDW}\sigma^3\tau^1\lambda^0.
\end{equation}where $\tau$ represent the Pauli matrix acting on the two components of $\Psi_{\tau\lambda}(\vec{k})$.

In the folded BZ in Fig.~\ref{fig:SDW}(b), the Fermi pockets appear in two separate sets. The `$\Gamma$-X' pockets result from the SDW folding of the pocket encircling the $X=(\pi,0)$ point onto the pocket encircling $\Gamma$, while the `Y-M' pockets result from the folding of the pocket encircling $M$ onto the pocket encircling the $Y=(0,\pi)$ point. Throughout the paper, we will focus on transverse momenta near the `$\Gamma$-X' pockets. The other Fermi pockets can be similarly considered.



 Since the SDW {Hamiltonian (\ref{hamiz}) commutes with $\sigma^3$, its eigenstates can be separated into two independent sectors.} 
And since the `$\Gamma$-X' pockets are constructed out of the bands $\mathcal{E}_{h}(\vec{k})$ and $\mathcal{E}_{e}(\vec{k}+\vec{Q})$, the states close to these pockets involve only the corresponding band-basis states, which we denote as $\varphi_{h,\vec{k}}$ and $\varphi_{e,\vec{k}+\vec{Q}}$, resp. 
Hence, an effective Hamiltonian for the low-energy spin-$\uparrow$ states can be written, 

\begin{equation}\mathcal{H}_{\uparrow\tau}(\vec{k}) =
\begin{bmatrix} \mathcal{E}_h(\vec{k}) & \Delta_{\scriptscriptstyle{SDW}}f(\vec{k}) \\  \Delta_{\scriptscriptstyle{SDW}}f(\vec{k}) & \mathcal{E}_e(\vec{k}+\vec{Q}) \end{bmatrix}, \ \  \psi_{\tau,\vec{k}} =\begin{pmatrix} \varphi_{h,\vec{k}} \\ \varphi_{e,\vec{k}+\vec{Q}} \end{pmatrix},
\end{equation}
where $f(\vec{k})= \braket{\varphi_{h,\vec{k}}| \varphi_{e,\vec{k}+\vec{Q}}}$ is the orbital overlap between states near the $\Gamma$ and $X$ pockets that are connected by the wavevector $\vec{Q}$.

Defining $E^{\pm}(k)= \frac{\mathcal{E}_h(k)\pm\mathcal{E}_e(k+Q)}{2}$, 
we can write the effective low-energy Hamiltonian for the SDW phase as
\begin{equation}\mathcal{H}_{\sigma\tau}(\vec{k}) = E^+(\vec{k})\tau^0 + E^-(\vec{k})\tau^3 + \Delta_{\scriptscriptstyle{SDW}}f(\vec{k})\sigma^3\tau^1.
\label{eq:noSC}\end{equation}

\begin{figure}[h]
\includegraphics[trim={5.1cm 9.3cm 5.1cm 9.3cm},clip,width=8.7cm,height=6.7cm]{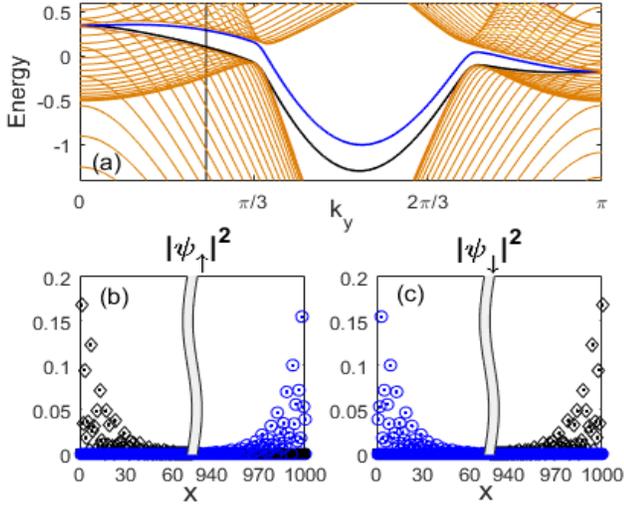}
\caption{(a): Spin-polarized edge bands in the bulk gap. (b)-(c): Amplitude of spin-resolved edge states (at $k_y$ shown by vertical line in (a)) corresponds to the upper (blue circles) and lower (black diamonds) bands in (a). In (a)-(c), $\Delta_{SDW}=.15$}
\label{fig:SDW-edge-bands}
\end{figure}

 With $(\pi,0)$-SDW the edge bands become spin-split (Fig.~\ref{fig:SDW-edge-bands} (a))\cite{Lau2013}. 
For a particular transverse momentum, subfigures (b)-(c) of Fig.~\ref{fig:SDW-edge-bands} display the relative amplitudes of each spin-sector for each of the separated edge bands.

\section{Superconducting SDW Phase}
\label{CoexistencePhase}

\begin{figure}[h]
\includegraphics[trim={5.2cm 9.3cm 4.8cm 9.4cm},clip,width=9cm]{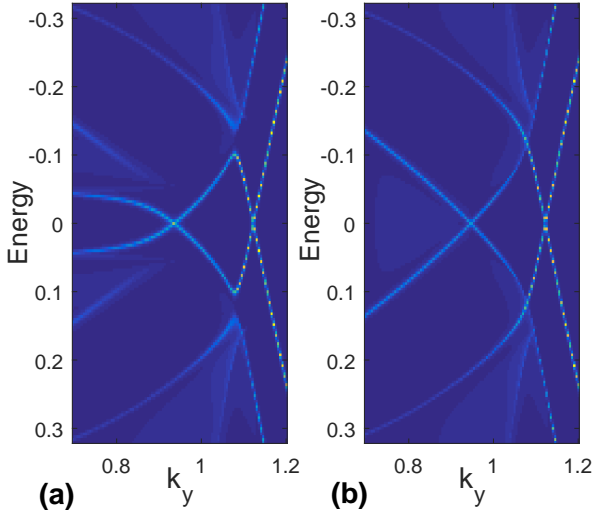}
\caption{Gapless edge bands in the SDW phases, (a): with ($\Delta_{\pm}^s=.05$), and (b): without ($\Delta_{\pm}^s=0$) superconductivity. In (a)-(b), $\Delta_{SDW}=.15$}
\label{fig:spectrals}
\end{figure}

{At the `$\Gamma$-X' pockets, the BdG Hamiltonian $
\mathcal{H}_{\mu\sigma\tau}(\vec{k})=\mu^3 \mathcal{H}_{\sigma\tau}(\vec{k})+\Delta_s^{\pm}\mu^2\sigma^2\tau^3$ commutes with $\mu^3\sigma^3$ } and its
eigenstates can be decoupled into two sectors corresponding to  $\braket{\mu^3\sigma^3}=\pm 1$. Each sector comprises two of the four eigenstates of $\mu^3\sigma^3$ and we define $\rho_i$ as the Pauli matrix acting on the two states in each of these independent two dimensional sectors.

In the coexistence phase of superconductivity and SDW the midgap bands shown in Fig.~\ref{fig:spectrals} are the particle-hole pair, with (\ref{fig:spectrals}(a)) and without (\ref{fig:spectrals}(b)) superconductivity corresponding to the spin-polarized edge bands near the Fermi level in Fig.~\ref{fig:SDW-edge-bands}. Fig.~\ref{fig:spectrals} shows the dispersion of edge states which corresponds to ABS that merge into the the topological edge bands near the first zero-energy crossing. This result is derived using an iterative surface Green's function method\cite{Lopez1985,Lopez1984} which does not suffer from hybridization of the states on the two edges.

\begin{figure}[h]
\includegraphics[trim={3.2cm 9.2cm 3.6cm 8.5cm},clip,width=9cm]{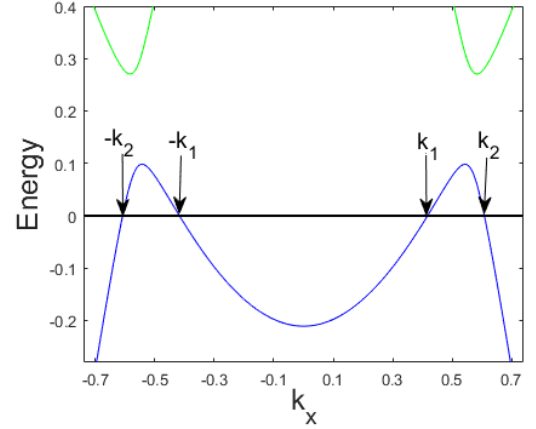}
\caption{Lower bulk SDW `$\Gamma$-X' band, and Fermi points $\pm k_{1,2}$, for $\Delta_{SDW}=.15$, and $k_y = .8$}
\label{fig:Fermi_points}
\end{figure}

To gain analytical insight into these edge states, we use the effective low energy Hamiltonian close to the Fermi points, for fixed $k_y$ momentum parallel to the edge\cite{Ghaemi2009}. Unlike in the unfolded case, where the low-energy bulk excitations of the normal phase for a given $k_y$ involves multiple bands, the four Fermi points in the $(\pi,0)$-folded case belong to the same band. Hence, for a given $k_y$, the low energy physics involves only one of the SDW-folded bands ploted in fig. (\ref{fig:Fermi_points}).

At the Fermi points, the linearized Hamiltonian is  
\begin{equation}\label{eq:HBdG}
\mathcal{H}^{F}_{\rho}(-i\partial_x,k_y)= -i\sum_{k_F} \rho^3v_F(k_F,k_y)\partial_x
+\Delta_{s}^\pm(k_F,k_y) \rho^1,
\end{equation}
where $v_F(k_F,k_y)$ is the Fermi velocity at each of the four Fermi points $k_F\in \{\pm k_{n=1,2}\}$ for a given $k_y$, as in Fig.~\ref{fig:Fermi_points}.



For a given transverse momentum, $k_y$, the edge states in the $\braket{\mu^3\sigma^3}=+1$ sector have the form $
\Psi_{\rho\lambda}^{\uparrow\text{edge}}(x,y)=\mathcal{N} e^{ik_y y}\Phi_{\rho\lambda}^{\uparrow F}(x),
$
where the ansatz $\Phi_{\sigma\lambda}^{\uparrow F}(x)$ is a superposition of states at the four Fermi points intersected by $k_y$. 
For a semi-infinite system, 
we construct these states to vanish at infinity as $\psi^{\uparrow n,\pm}_{\rho\lambda}(x) = e^{-\frac{x}{\lambda_n}}\phi^{\uparrow n,\pm}_{\rho\lambda}$, where
each $\phi^{\uparrow n,\pm}_{\rho\lambda}$ is a Nambu-spinor containing the appropriate orbital content, and each $\lambda_n$ represents a decay length. In terms of the BdG coherence factors, 

\begin{equation}
\phi^{\uparrow n,\pm}_{\rho\lambda}=\begin{pmatrix}
u_{\uparrow,xz}^{n,\pm}\\ u_{\uparrow,yz}^{n,\pm} \\ \nu_{\downarrow,xz}^{n,\pm} \\ \nu_{\downarrow,yz}^{n,\pm}
\end{pmatrix}=\begin{pmatrix}
\Delta(k_n) B(\pm k_n)\varPsi^{e\uparrow}_{\lambda}(k_n)\\ 
\mathcal{E}(\pm k_n) B(\pm k_n)\varPsi^{h\downarrow}_{\lambda}(k_n)
\end{pmatrix},
\label{NambuSpinorBasis}
\end{equation}
where $\Delta(k_n) = \Delta_s^\pm(k_n;k_y)$ and $\mathcal{E}(\pm k_n)= E+\frac{v_F(\pm k_n,k_y)}{i\lambda_n}=E\pm (-1)^n \sqrt{E^2 - \Delta(k_n)^2}$ . Here $E$ is the energy of the edge state, $\varPsi^{e\uparrow}_{\lambda}$ and $\varPsi^{h\downarrow}_{\lambda}$ are the eigenvectors corresponding to the $\braket{\rho^3}=+1$ and $\braket{\rho^3}=-1$ diagonal blocks of bulk superconducting Hamiltonian, and  
\begin{equation}
B(k_n)=\left[\begin{array}{cc}
r_2(k_n) & r_1(k_n+\pi) \\
s_2(k_n) & s_1(k_n+\pi)
\end{array}\right],
\end{equation} 
where $\left(r_1(k_n),s_1(k_n)\right)^T$ and $\left(r_2(k_n),s_2(k_n)\right)^T$ are the orbital wave functions of the states on electron and hole bands resulting from the two orbital Hamiltonian given in (\ref{OrbHam}). 
The complete edge wave function can be written in the form

\begin{equation}
\Psi^{\uparrow \text{edge}}_{\rho\lambda}(x,y)= e^{ik_y y}\sum_{\substack{n=1,2 \\ m=-1,1}}C_{nm}e^{mi k_n x} \psi^{\uparrow n m}_{\rho\lambda}(x),
\label{BoundaryWaveFunc}
\end{equation}
where $m=+(-)1$ labels the Fermi points to the right (left) of the origin. The boundary condition at the edge, $\Psi^{\uparrow \text{edge}}_{\rho\lambda}(0,y)=0$, implies that $\det{\begin{pmatrix}
\phi^{\uparrow 1,+}_{\rho\lambda} & \phi^{\uparrow 2,+}_{\rho\lambda} &
\phi^{\uparrow 1,-}_{\rho\lambda} & \phi^{\uparrow 2,-}_{\rho\lambda}
\end{pmatrix}}=0$ which gives the relation between decay lengths, energy, and bulk parameters.

\begin{figure}[h]
\includegraphics[trim={5.9cm 10cm 5.6cm 9.9cm},clip,width=8.4cm,height=7cm]{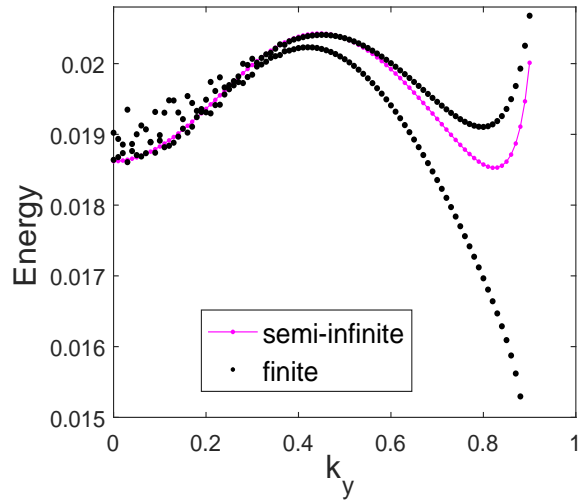}
\caption{Edge dispersion for semi-infinite system (magenta), compared with spin-slit edge bands (black) from exact diagonalization. $\Delta_{SDW}=.15$, and $\Delta_{\pm}^s=.04$.}
\label{fig:Disps-semiinf_vs_fin}
\end{figure}

Figure \ref{fig:Disps-semiinf_vs_fin} shows the edge state dispersion (magenta curve) obtained for a semi-infinite system through the semi-analytic treatment outlined above, compared with the spin-slit edge bands (black dotted curves) obtained through exact diagonalization of the finite lattice model. 
The spin polarization at the edge appears naturally because of the explicit specification of the magnetization at each site in the antiferromagnetic lattice Hamiltonian. On the other hand, our treatment of the semi-infinite case is in the continuum limit, where there is no such specification for the doubled unit cell. We should note that the semi-analytical treatment is only valid for the states close to the bulk Fermi surface. This is also apparent in Fig.~\ref{fig:Disps-semiinf_vs_fin} where these results deviate from the exact diagonalization results for transverse momenta where the bulk Fermi surfaces vanish in the normal phase.
The range of momentum $k_y$ where the bulk contains the gapless states at the Fermi surfaces is particularly relevant for the proximity induced odd-frequency pairing studied in the next section. Our semi-analytical treatment, which does not suffer from finite size effects, particularly supports the applicability of our results to real pnictide materials. 

\section{Odd-frequency pairing}
\label{OddFrequency}


In order to investigate the induced odd-frequency pairings, we  calculate the two-fermion anomalous correlators\cite{Ebisu2015,BlackSchaffer2013,BlackSchaffer2013Jun,Asano2013,Higashitani2014} of electrons with different time coordinates at locations $x_i$ and $x_j$, \cite{Abrahams1995,Linder2017}.
\begin{equation}
\tilde{\mathcal{F}}_{t}(i\sigma,j\sigma^\prime)
= -\braket{T\psi^\dagger_{\sigma}(x_i,t)\psi^\dagger_{\sigma^\prime}(x_j,0)},
\end{equation}
where $T$ is the time-ordering operator. In the Matsubara representation\cite{Ebisu2015}, 
\begin{equation}
\tilde{F}_\omega(i\sigma,j\sigma^\prime)
=\sum_{n}^\prime \Big(\frac{\bar{u}_{n\sigma i}\nu_{n\sigma^\prime j}}{i\omega+E_n}+\frac{\bar{u}_{n\sigma^\prime j}\nu_{n\sigma i}}{i\omega-E_n}\Big),
\end{equation}
where $E_n$ are the positive eigenenergies of the BdG Hamiltonian \ref{eq:HBdG} and $u_{n\sigma i}(v_{n\sigma i})$ are the spin-$\sigma$ electron (hole) components of the corresponding eigenvectors at position $x_i$, obtained self-consistently by solving the BdG equation with the position-dependent order-parameter\cite{Eschrig2015}
\begin{equation}
 \Delta_{i\sigma,i+1\sigma^\prime}=g\big(\Delta^s (i\sigma^2)_{i\sigma,i+1\sigma^\prime} +\vec{\Delta^t}\cdot (i\vec{\sigma}\sigma^2)_{i\sigma,i+1\sigma^\prime}\big),
\end{equation}
where $\Delta^s$ and $\vec{\Delta^t}$ denote the singlet and triplet components of the equal time correlator, $\mathcal{F}_0(i\sigma,i+1\sigma^\prime)$, and g is the nearest-neighbor interaction potential.


\begin{figure}[h]
\hspace*{-0.5cm}
\includegraphics[trim={2.0cm 6cm 2.4cm 6cm},clip, width=9.1cm,height=6.2cm]{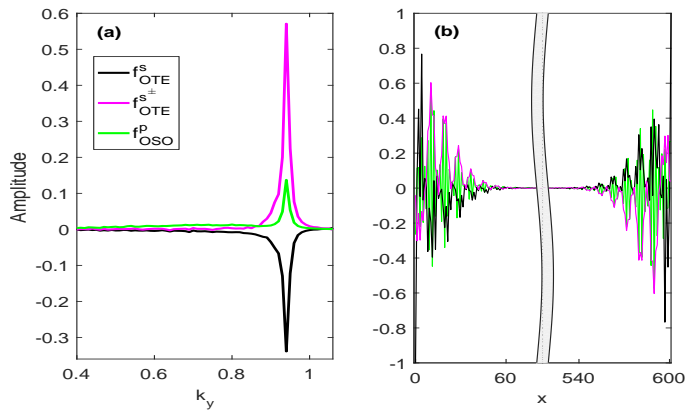}
\caption{Odd-frequency pairing: OSO p-wave (green), OTE s-wave (black), and OTE $s^\pm$-wave (magenta), for a 600-site lattice with $\Delta_{SDW}=.15$ and $\Delta_{\pm}^s=.04$. (a) shows the odd-frequency amplitudes (normalized by the bulk even-frequency s-wave amplitude) as a functions of transverse momentum $k_y$. (b) Decay of odd-frequency pairing amplitude into the bulk (normalized by their maximum value at the edge).}
\label{fig:pairingsky}
\end{figure}

Pairing amplitudes are then constructed as antisymmetric linear superpositions of anomalous correlators. Even (odd) frequency amplitudes must  be manifestly antisymmetric (symmetric) in the coordinates other than frequency. The dominant odd-frequency intra-orbital pairings generated in our setup belong to two classes: spin-singlet, odd-parity (OSO) and spin-triplet, even-parity (OTE). The relevant odd-frequency s-wave and extended s-wave pairing amplitudes are,
\begin{multline}
\label{eq:OTEsw}
f_{OTE}^{s} = \frac{1}{2}(\tilde{F}_\omega(i\sigma ,i\sigma^\prime)+\tilde{F}_\omega(i\sigma^\prime ,i\sigma)) \\ =\frac{\omega}{i}\sum_n^\prime \frac{\bar{u}_{n\sigma i}\nu_{n\sigma^\prime i}+\bar{u}_{n\sigma^\prime i}\nu_{n\sigma i}}{\omega^2 + E_n^2},
\end{multline}
and
\begin{multline}
\label{eq:OTExsw}
f_{OTE}^{s_{\pm}} = \frac{1}{4}(\tilde{F}_\omega(i+1 \ \sigma ,i\sigma^\prime)+\tilde{F}_\omega(i\sigma^\prime ,i+1 \ \sigma) \\
+\tilde{F}_\omega(i\sigma ,i+1 \ \sigma^\prime)+\tilde{F}_\omega(i+1 \ \sigma^\prime ,i\sigma))\\ =\frac{\omega}{2i}\sum_n^\prime \frac{1}{\omega^2 + E_n^2}(\bar{u}_{n\sigma i+1}\nu_{n\sigma^\prime i}+\bar{u}_{n\sigma^\prime i}\nu_{n\sigma i+1} \\ + \bar{u}_{n\sigma i}\nu_{n\sigma^\prime i+1}+\bar{u}_{n\sigma^\prime i+1}\nu_{n\sigma i} ),
\end{multline}
resp., while the relevant odd-frequency p-wave pairing amplitude is
\begin{multline}
\label{eq:OSOpw}
f_{OSO}^{p} = \frac{1}{4}(\tilde{F}_\omega(i+1 \ \sigma ,i\sigma^\prime)+\tilde{F}_\omega(i\sigma^\prime ,i+1 \ \sigma) \\
-\tilde{F}_\omega(i\sigma ,i+1 \ \sigma^\prime)-\tilde{F}_\omega(i+1 \ \sigma^\prime ,i\sigma))\\ =\frac{\omega}{2i}\sum_n^\prime \frac{1}{\omega^2 + E_n^2}(\bar{u}_{n\sigma i+1}\nu_{n\sigma^\prime i}+\bar{u}_{n\sigma^\prime i}\nu_{n\sigma i+1} \\ - \bar{u}_{n\sigma i}\nu_{n\sigma^\prime i+1}-\bar{u}_{n\sigma^\prime i+1}\nu_{n\sigma i} ).
\end{multline}

The breaking of translational symmetry at the edge of the extended s-wave superconductor generates surface ABS with p-wave singlet odd-frequency pairing, $f_{OSO}^{p}(x_i,k_y)$, for both the paramagnetic and antiferromagnetic (green curves in Fig.~\ref{fig:pairingsky}) phases. In the coexistence phase of SDW and extended s-wave superconductivity, due to the breaking of time-reversal and translational symmetries, edges further accommodate odd-frequency spin-triplet s-wave, $f_{OTE}^{s}(x_i,k_y)$, and extended s-wave, $f_{OTE}^{s\pm}(x_i,k_y)$, pairings. The enhancements of these near edges can be seen in Fig.~\ref{fig:pairingsky}(b) (black and magenta curves, resp.). As was noted in previous studies\cite{Ghaemi2011,Hinojosa2014}, breaking of time-reversal symmetry by SDW in the coexistence phase induces triplet (even-frequency) pairings, which dominate in certain regions of the BZ (near the tips of the the folded pockets in Fig.~\ref{fig:SDW}(b)). The translational symmetry breaking due to the edge then generates odd-frequency s-wave triplet components.


Since the energies of the edge states appear in the denominators of (\ref{eq:OTEsw}-\ref{eq:OSOpw}), odd-frequency pairing amplitudes are enhanced near the midgap crossings. Hence, emergent odd-frequency pairings of the types considered here are optimized within parameter ranges that yield midgap edge modes as close as possible to the tips of the bulk Fermi pockets, as in Figs.~\ref{fig:pairingsky}(a)-(c).

\begin{figure}[h]
\vspace{-2.5cm}
\hspace{0cm}
\includegraphics[width=8.4cm]{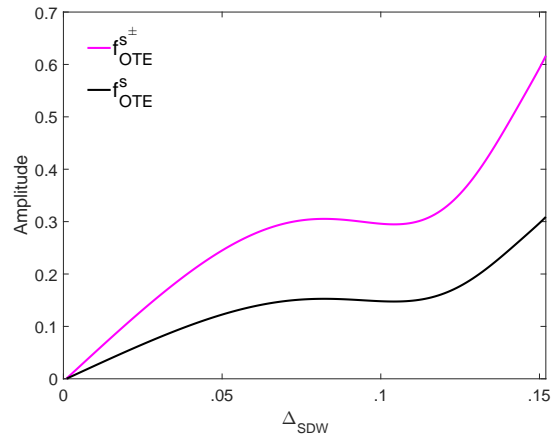}
\vspace{-2.5cm}
\caption{Odd-frequency triplet pairing amplitudes as a function of $\Delta_{SDW}$, for $\mu_F=1.1$, $k_y=.91$, $\Delta_{\pm}^s=.04$ (normalized by the bulk even-frequency s-wave amplitudes).}
\label{fig:OTEvsSDW}
\end{figure}

In the limit of vanishing magnetism, spin-rotational symmetry is regained and the odd-frequency triplet pairing is suppressed (see Fig. \ref{fig:OTEvsSDW}). 



The dependence of odd-frequency triplet pairing on the coexistence of superconductivity and SDW can also be understood from the structure of the corresponding anomalous correlators. The spin-$\downarrow$ sector analogue of our edge-state solution (\ref{BoundaryWaveFunc}) is obtained by replacing $\Delta_{\pm}^s$ with $-\Delta_{\pm}^s$ and $\Delta_{SDW}$ with $-\Delta_{SDW}$. 
When SDW vanishes, it can be shown from (\ref{NambuSpinorBasis}) that the coherence factors corresponding to the $\uparrow$ and $\downarrow$ sectors are related as $u^{nm}_{\downarrow \lambda}=-u^{nm}_{\uparrow \lambda}$ and $\nu^{nm}_{\downarrow \lambda}=\nu^{nm}_{\uparrow \lambda}$, which implies the vanishing of the anomalous correlators given in (\ref{eq:OTEsw}) and (\ref{eq:OTExsw}).

\section{Five-orbital model and effect of spin-orbit coupling}\label{fiveo}
\begin{figure}
\includegraphics[width=8.5cm]{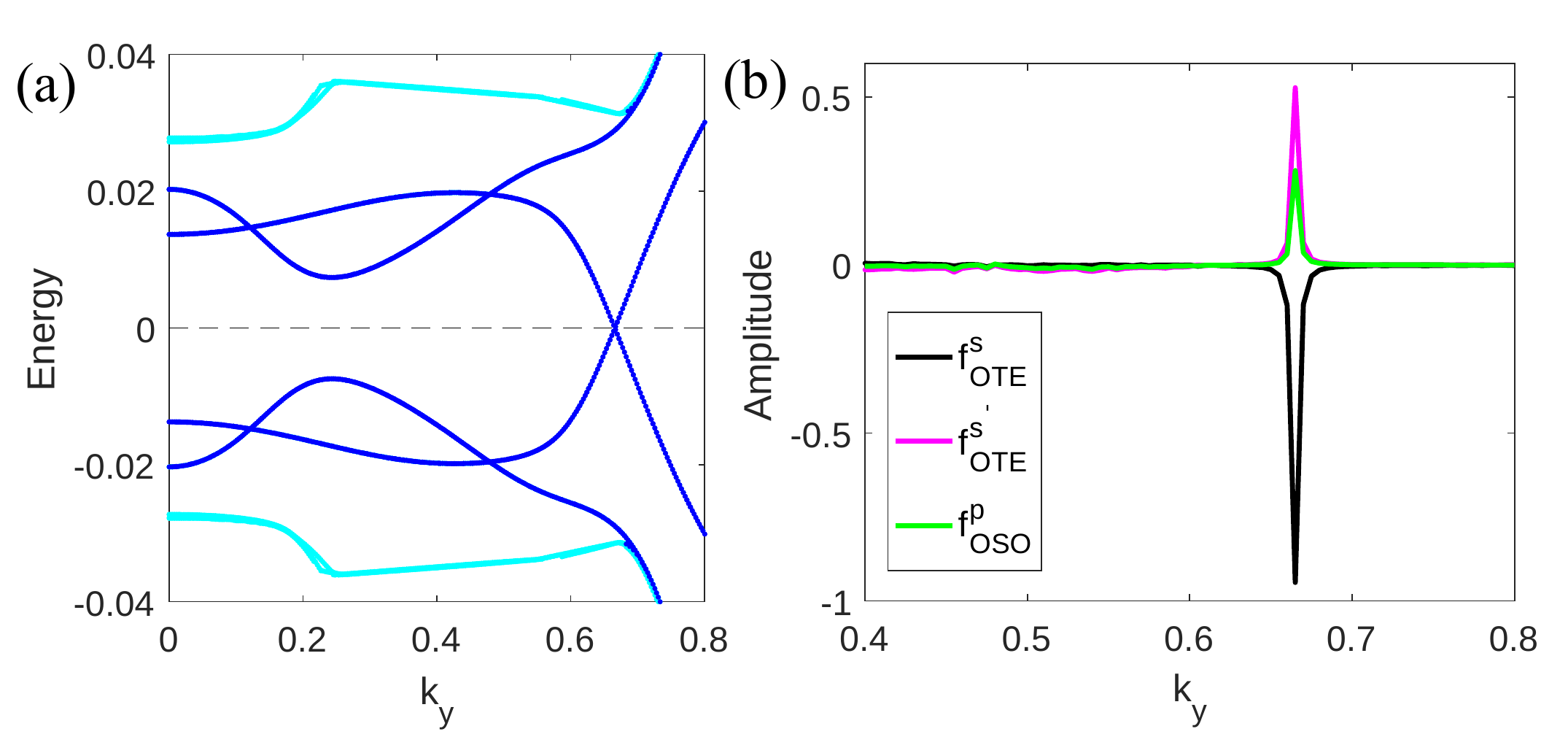}
\caption{(a) Dependence of surface bands (blue) and low lying bulk bands (cyan) of five orbital model on transverse momentum $k_y$. (b) Odd-frequency pairing amplitudes: OSO p-wave (green), OTE s-wave (black), and OTE $s^\pm$-wave (magenta). Results are for 300-site lattice with $\Delta_{SDW}=0.14$ and $\Delta_{\pm}^s=0.04$.}
\label{fig:FiveOrbNoSOC}
\end{figure}

\begin{figure}[h]
\hspace*{-0.5cm}
\includegraphics[trim={2.6cm 7.7cm 4.1cm 8.1cm},clip, width=7.7cm,height=4.8cm]{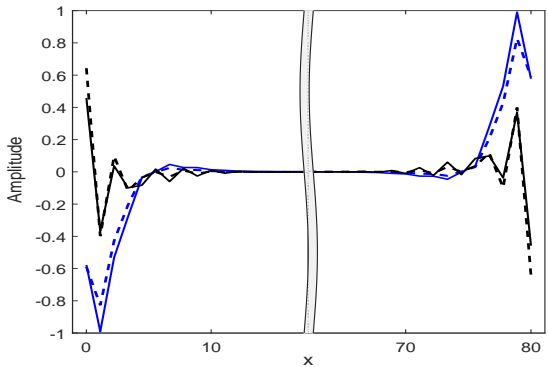}
\caption{Longitudinal profiles of OTE $s^\pm$-wave (blue) and OSO p-wave (black) amplitudes at $k_y=.671$, obtained before (dashed) and after (solid) self-consistent calculation of the gap function at t=0, for an 80-site lattice with $\Delta_{SDW}=.14$, and $\Delta_{\pm}^s=.04$ (normalized by max. value at the edge).}
\label{fig:selfcons}
\end{figure}

It is well-known that the two orbital model undermines some of the crucial properties of the pnictides \cite{Cvetkovic2013,Fernandes2017}. In particular, the two orbital model does not respect the glide-plane symmetry of the lattice. Since the symmetries of the parent materials constrain the types of superconductivity that can form, it is important to determine whether the artificial breaking of lattice symmetries in the two-orbital model has qualitative consequence on the obtained results. Hence, in this section we present our results on the emergence of odd-frequency pairing using the more elaborate five orbital model \cite{Graser2009}, with an effective 10-band tight-binding description in order to account for the 2-Fe unit-cell, which becomes necessary when considering spin-orbit coupling\cite{Fernandes2017}. Due to the complexity of this model, we employed an exact diagonalization method on a finite lattice. Fig.~\ref{fig:FiveOrbNoSOC} (a) shows the resulting surface bands and low lying bulk bands for
this model. As can be seen from the figure we obtain similar Andreev surface bands as in the two orbital model. Corresponding odd-frequency correlators are presented in Fig.~\ref{fig:FiveOrbNoSOC} (b). As in the two orbital model, we observe both OTE and OSO pairings, with the OTE s-wave magnitudes dominating. As is clear from the forms of their constituent correlators, the odd-frequency amplitudes are peaked at the positions of the zero energy band crossings, where the even frequency components vanish. Figure \ref{fig:selfcons} shows the spatial profile of the OTE extended s-wave and OSO p-wave amplitudes at the momentum corresponding to the crossing in figure \ref{fig:FiveOrbNoSOC} (a), as obtained with and without self-consistent treatment of the equal-time superconducting order parameter (solid and dashed lines, resp.). These results demonstrate that both the occurrence of Andreev surface states and the induced odd-frequency correlators do not rely on specific model of the band structure and should be detectable in experimental settings.

\begin{figure}
\includegraphics[width=8.6cm]{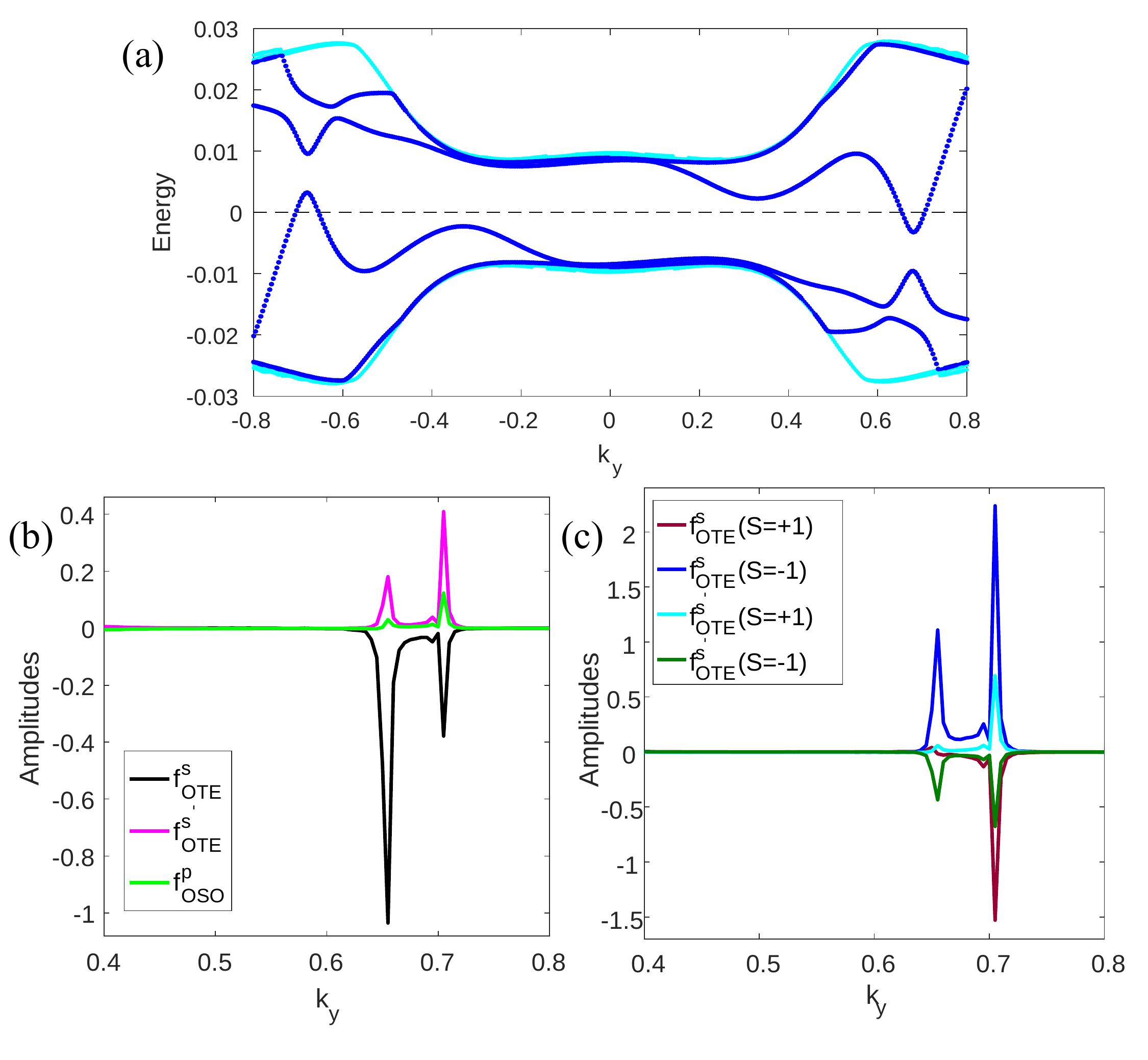}
\caption{Same as in Fig.~\ref{fig:FiveOrbNoSOC}, with SOC $\Delta_\mathrm{SOC}=0.06$ added.
 (c) Odd-frequency amplitudes for spinful $S=\pm1$ cases with SOC.}
\label{fig:FiveOrbSOC}
\end{figure}

Another advantage of the five-orbital tight-binding model is that it allows for the inclusion of spin-orbit coupling (SOC) in the Hamiltonian. It has been shown that the presence of SOC in pnictides has important effects on their properties \cite{Borisenko2016}. Doubling to the 2-Fe unit cell and including SOC results in hybridization between the $X$ and $Y$ pockets\cite{Christensen2015}. Fig.~\ref{fig:FiveOrbSOC} (a) shows the resulting surface and low-lying bulk bands. The inclusion of SOC opens a gap at each crossing of spin-polarized edge states, due to the spin mixing. Also note that at each $k_y$ the spectrum is not symmetric with respect to zero energy. In this case particle-hole symmetry connects states at $k_y$  and $-k_y$ with opposite energies.
 SOC also has important consequences for odd-frequency correlators, as depicted in Fig.~\ref{fig:FiveOrbSOC} (b,c). First, the spin zero correlators now have a double peak structure, corresponding to the two zero crossings of the SOC-gapped surface band. In addition to that we observe the emergence of spinful ($S=\pm1$) triplet odd frequency components (see Fig.~\ref{fig:FiveOrbSOC} (c)). When SOC is absent the BdG Hamiltonian has a rotational symmetry around the SDW quantization axis, which forces the correlators $\tilde{F}(i\sigma,j\sigma)$ to vanish. SOC breaks this symmetry, thereby facilitating the generation of odd-frequency triplet correlators between states with the same spin component along the axis of spin-quantization for the case without SOC.    


\section{Discussions and conclusion}
\label{Conclusion}

By now, many proposals have been put forward to detect signatures of odd-frequency pairing in different systems. Before proceeding to actual proposals for observing odd-frequency pairing in the current system, we should note that obtaining non-zero odd-frequency correlators is directly linked with the presence of in-gap Andreev surface states. As noted in several works \cite{Ghaemi2009,OnariTanaka2009,Nagai2009} the in-gap Andreev surface states are not possible for s-wave superconductivity and they are a direct consequence of extended s-wave pairing. In addition, the appearance of OTE relies on breaking of spin-rotational symmetry, which is due to the presence of SDW. In other words the observation of signatures of OTE would indirectly verify the extended s-wave structure and its coexistence with SDW in pnictides.

From the numerous proposals for experimental verification of odd-frequency pairing, such as the modification of the density of states in diffusive ferromagnet/superconductor junctions \cite{{Yokoyama2007,Tanaka2007Jan}}, the paramagnetic Meissner effect \cite{Yokoyama2011,Komendova2015,DiBernardo2015Meissner}, and non-local transport signatures due to crossed Andreev reflection in topological insulators \cite{Crepin2015,Beiranvand2017,Fleckenstein2018}, the ones with the closest relevance to our system are those involving superconductor/magnetic-interface/normal metal heterostructures \cite{Linder2009}. The proximity effect between superconductor and normal metal results in the development of a gap in the metal band structure. It has also been shown that odd-frequency pairing would lead to an enhancement of the density of states at zero energy, which can be detected in scanning tunneling microscopy (STM) measurement. Our results show that OTE pairing only emerges at edges of pnictides when SDW coexists with superconductivity (see figure \ref{fig:OTEvsSDW}). Taking into account that such pairing is robust in the diffusive regime, placing a pnictide sample next to a metal should induce a similar modification of its density of states. This can also be detected in STM measurements. In addition to this, the coexistence of SDW and superconductivity can be controlled by changing temperature and doping level. A pnictide sample can be tuned from the superconducting phase without SDW into the phase where SDW and superconductivity coexist. This will result in an increase of the superconducting gap in the metal, and at the same time, the presence of OTE will lead to the enhancement of the density of state at zero energy, which was absent in the phase without SDW. Therefore, pnictides are naturally suited to control and observe signatures of odd-frequency pairing. This can lead to unambiguous detection of odd-frequency superconducting pairing in future experiments.

In conclusion, we have studied the edge states of pnictide superconductors in the coexistence phase of stripe antiferromagnetic SDW and extended s-wave superconductivity and have shown, self-consistently, that the edge states can develop odd-frequency superconducting pairing. Without SDW, while explicit translational symmetry breaking can in principle lead to odd-frequency pairing at edges, it can not induce triplet pairs, because it does not break spin-rotational symmetry. Thus any odd-frequency pairing at edges in the paramagnetic phase would have to be odd under parity, and hence would not be robust to disorder. However, when SDW is also present, the additional breaking of spin-rotational symmetry further generates odd-frequency triplet pairing amplitudes, which are even under parity and would thus persist in a more realistic, diffusive regime.

The model we considered predicts the emergence of odd-frequency triplet pairing on edges of pnictides in the coexistence phase of SDW and extended s-wave superconductivity. We also showed that the inclusion of SOC\cite{Cvetkovic2013,Vafek2017} generates an odd-frequency triplet pairing between states with the same spin. In addition, we believe that our results could shed light on the apparent enhancement of superfluid density at defects in the SDW ordering in pnictide superconductors. The detailed study of such effects will be a subject of our future studies.


\section*{acknowlegments}
We would like to thank Rafael Fernandes, Sarang Gopalakrishnan, Parameswaran Nair, Alexios Polychronakos and Javad Shabani for useful discussions. This work was supported in part by the National Science Foundation CREST Center for Interface Design and Engineered Assembly of Low Dimensional systems (IDEALS), NSF grant number HRD-1547830 (CY) and NSF Grant No. EFRI-154286 (AG and PG).

\bibliography{bibpnictides}

\end{document}